\documentclass{PoS}

\title{Chiral extrapolation of matrix elements of BSM kaon operators}

\ShortTitle{Chiral extrapolation of matrix elements of BSM kaon operators}

\author{Jon A. Bailey\\
        Department of Physics and Astronomy, Seoul National University, 
        Seoul, 151-747, South Korea} 
\author{Hyung-Jin Kim\\
        Physics Department, Brookhaven National Laboratory, Upton, 
        New York 11973, USA} 
\author{Weonjong Lee\\
        Department of Physics and Astronomy, Seoul National University, 
        Seoul, 151-747, South Korea} 
\author{\speaker{Stephen R. Sharpe}\\%
        Physics Department University of Washington Seattle, 
        WA 98195-1560, USA\\
        E-mail: \email{sharpe@phys.washington.edu}}

\abstract{Models of new physics induce $K_0-\overline{K_0}$ mixing 
through operators having Dirac structures other than the ``left-left'' 
form of the Standard Model. 
To carry out the chiral-continuum
extrapolation of results from numerical simulations, 
one needs to know the quark mass and lattice spacing dependence of the
corresponding B-parameters in the partially quenched theory at least at
next-to-leading order. For simulations using staggered fermions
(such as that we are doing with HYP-smeared valence fermions on the
MILC asqtad lattices)
one must determine this dependence using staggered chiral perturbation
theory (SChPT).
We have calculated the required dependence in both SU(3) and SU(2) SChPT, 
working at next-to-leading order,
and we give here an overview of the methodology and results.
The SU(3) SChPT result turns out to be
much simpler than that for the Standard Model $B_K$ operator, due to the
absence of chiral suppression for the new operators. The SU(2) SChPT
result turns out to be closely related to that for $B_K$:
the chiral logarithms are identical, up to an operator-dependent sign.
Our results are
also useful for fermions with chiral symmetry as they provide, in the
continuum limit, the partially quenched generalization of existing continuum
results.}

\FullConference{The 30th International Symposium on Lattice Field Theory\\
		 June 24-29,  2012\\
		 Cairns, Australia}

\begin{document}

\section{Introduction}

Theories of physics beyond the standard model (BSM) are often
highly constrained by the observed weakness of flavor-changing
neutral processes. One of the strongest constraints comes from
$K_0-\overline{K_0}$ mixing, both the CP-conserving and violating parts.
For a given model of new physics, integrating out heavy particles leads
to a set of local $\Delta S=2$ operators with known Wilson coefficients.
In order to determine the constraint placed on the parameters of the model,
one must calculate the $K_0-\overline{K_0}$ 
matrix elements of the $\Delta S=2$ operators.
This is a task for which lattice methods are well suited, 
and several efforts are underway~\cite{dwf_bsm_pap,tm_bsm_pap,stag_bsm_proc}.
The results will provide complementary information to the direct searches
for new physics at the LHC.\footnote{%
For a recent review of this complementary information using the first
lattice results see Ref.~\cite{mescia12}.}

Discretization effects and unphysically large $u$ and $d$ quark
masses are important sources of systematic error in these
calculations, which fortunately can be largely eliminated by
performing chiral and continuum extrapolations with effective field
theory as a guide.  The need for controlled, model-independent
continuum extrapolations is particularly acute for our calculation
because we use rooted staggered quarks, and one must extrapolate away
the effects of taste-breaking and rooting.
To do this, one requires at least a next-to-leading order (NLO) calculation
in staggered chiral perturbation theory (SChPT).
Here we report on such a calculation, which has been
recently published~\cite{bsmprd}.
We have presented results for both SU(3) and SU(2) SChPT, i.e. treating
the strange quark as light and heavy, respectively.
Our results are also useful for describing data obtained with other fermion
discretizations, e.g. domain-wall fermions~\cite{dwf_bsm_pap}, for
we can turn off taste breaking and determine the NLO predictions
for a partially quenched (PQ) continuum theory in both SU(3) and SU(2) cases.

Since all the technical details have been provided in Ref.~\cite{bsmprd},
we attempt here a complementary discussion, focusing on the essential
features and eschewing most details. We also to correct some minor
shortcomings in the original discussion, which, however,
lead to no changes in the final results.

BSM theories lead to $\Delta S=2$ operators with Dirac structures
which are not constrained to have the ``left-left'' structure of
the standard model (SM) operator. We adopt the basis
\begin{eqnarray}
 O_1 &=&
 [\bar{s}^a \gamma_\mu (1-\gamma_5) d^a] 
 [\bar{s}^b \gamma_\mu (1-\gamma_5) d^b] \,,
\label{eq:O1Cont}
\\
 O_2 &=&
 [\bar{s}^a (1-\gamma_5) d^a] 
 [\bar{s}^b (1-\gamma_5) d^b] \,,
\label{eq:O2Cont}
\\
 O_3 &=&
 [\bar{s}^a \sigma_{\mu\nu} (1-\gamma_5) d^a] 
 [\bar{s}^b \sigma_{\mu\nu} (1-\gamma_5) d^b] \,,
\label{eq:O3Cont}
\\
 O_4 &=&
 [\bar{s}^a (1-\gamma_5) d^a] 
 [\bar{s}^b (1+\gamma_5) d^b] \,,
\label{eq:O4Cont}
\\
 O_5 &=&
 [\bar{s}^a \gamma_\mu (1-\gamma_5) d^a] 
 [\bar{s}^b \gamma_\mu (1+\gamma_5) d^b] \,,
\label{eq:O5Cont}
\end{eqnarray}
where $\sigma_{\mu\nu}=[\gamma_\mu,\gamma_\nu]/2$, and $a$ and $b$ are
color indices.
This basis was used in the highest-order continuum calculations of
anomalous dimensions~\cite{buras}, 
and is the most natural choice
for our staggered calculation of the matrix elements, in which
we match onto continuum operators using 1-loop perturbation theory 
(PT)~\cite{Kim:2011pz}.

Many authors, including Refs.~\cite{dwf_bsm_pap,tm_bsm_pap,bsmprd},
use an alternate basis in which $O_3$ and $O_5$ are replaced by
\begin{eqnarray}
 O'_3 &=&
 [\bar{s}^a (1-\gamma_5) d^b] 
 [\bar{s}^b (1-\gamma_5) d^a] \,,
\label{eq:Op3Cont}
\\
 O'_5 &=&
 [\bar{s}^a (1-\gamma_5) d^b] 
 [\bar{s}^b (1+\gamma_5) d^a] \,,
\label{eq:Op5Cont}
\end{eqnarray}
We do not discuss the pros and cons of the two bases here, since 
the choice of basis turns out to be irrelevant for the SChPT calculation
as {\em the chiral behavior of the matrix elements of
$O_3$ and $O'_3$ (and of $O_5$ and $O'_5$) are identical.}
Thus the results presented here hold for either basis.

In the SM, only the operator $O_1$ appears, its matrix element being
parametrized by $B_K$. A large effort by the lattice community
has led to an accurate determination of $B_K$. 
The matrix elements of the other operators, 
which we call ``BSM operators'',
are in fact {\em easier} to calculate on the lattice. This is
because they do not vanish in the SU(3) chiral limit, and so have
simpler chiral expansions. Thus lattice results for the new matrix elements
are following closely on the heels of those for $B_K$.

As for $B_K$, calculating ratios of matrix elements is advantageous 
because many lattice systematics and
also some of the chiral logarithms cancel. 
Standard ratios for the BSM operators are
\begin{eqnarray}
 B_j (\mu) &=& 
\frac{\langle \overline{K}_0 | O_j(\mu) | K_0 \rangle}
     {N_j \langle \overline{K}_0 | \bar{s}^a \gamma_5 d^a(\mu)|0\rangle
      \langle 0 | \bar{s}^b \gamma_5 d^b(\mu)|K_0\rangle}\,,
\quad (N_2,N_3,N_4,N_5) = (5/3,4,-2,4/3)\,, 
\label{eq:BjNj}
\end{eqnarray}
where $\mu$ is the renormalization scale.
Note that, unlike for $B_K$, the denominator does not vanish in the
SU(3) chiral limit. 
Below we present and discuss our results for the 
chiral-continuum behavior of these $B_j$.

\section{Overview of methodology}

Our numerical calculations use a mixed-action set-up
with  HYP-smeared valence staggered quarks
atop $2+1$ flavors of rooted asqtad sea quarks (MILC configurations).
We allow the valence $d$ and $s$ masses, labeled $m_x$ and $m_y$,
respectively, to differ from the sea-quark masses, labeled
$m_u$, $m_d$ and $m_s$. Although our present simulations have $m_u=m_d$,
we consider the completely non-degenerate case in the SChPT 
calculation.\footnote{%
All masses are in physical rather than lattice units in this article.}

The calculation largely follows the methodology worked out 
in Ref.~\cite{VdWSS} for $B_K$ in PQ SChPT 
and extended to  mixed action SChPT in Ref.~\cite{bkprd}.
In most respects the calculation for the BSM operators
is more straightforward than that for $B_K$ because
the matrix elements do not vanish in the SU(3) chiral limit.
There is, however, an additional complication whose origin is
the non-trivial mixing between the BSM operators in PT.

In a continuum ChPT calculation, one would proceed by first mapping the
continuum BSM operators into chiral operators in the effective theory, and
then calculating one-loop diagrams. The operator mapping 
is done using transformation properties
under the chiral group $SU(N_f)_L\times SU(N_f)_R$.
To carry this over to staggered ChPT one must deal with
several complications:
\begin{itemize}
\item
The presence of additional valence tastes implies that the lattice
operators are, even in the continuum limit, in a partially quenched theory
with an enlarged chiral group. 
Because Fierz rearrangement differs in the presence of tastes,
one must introduce
two down and two strange valence fields ($D_1$, $D_2$, $S_1$ and $S_2$,
where uppercase indicates a field with four tastes)
in order to have a theory in which Wick contractions can be matched
with those of the desired continuum theory~\cite{VdWSS}.
One must also choose the taste of the lattice operator---with the
standard choice being the Goldstone taste $\xi_5$.
\item
Taste-breaking in the staggered action away from the continuum limit
introduces additional terms in the chiral Lagrangian which
are proportional to $a^2$, and, in the standard power-counting,
these terms appear at leading order (LO) 
along with terms proportional to $m$ and $p^2$~\cite{LS_SChPT}.
\item
The sea quarks are rooted and differ from the valence quarks.
The rooting can be dealt with in SChPT using the replica method of
Ref.~\cite{AB_SChPT}. The mixed action introduces further 
complications~\cite{bkprd} that, however, turn out to be unimportant
for the BSM operators (though not for $B_K$).
\item
Taste-breaking due either to the staggered action or introduced by
the use of inexact (i.e. one-loop) matching leads to many new
chiral operators. This proliferation of operators
is a major problem for $B_K$ [in SU(3) ChPT]
but turns out to be a minor issue here as the new operators only contribute
to analytic terms at NLO in SChPT (and not to loops).
\end{itemize}
To deal with these complications one must follow a rather elaborate
series of matching steps. The details are given in Ref.~\cite{bsmprd}.
Here we give only an overview.\footnote{%
This discussion extends and corrects that of Ref.~\cite{bsmprd},
since we have subsequently found that additional matching steps
(steps 1 and 2 below)
are required. The final results are, however, unaffected.
Details will be presented in Ref.~\cite{inprogress}.}

We illustrate the required steps by focusing on $O_2$.
In {\bf step 1}, we match from QCD to a PQ continuum theory 
containing two valence down quarks, $d_{1,2}$,
and two valence strange quarks, $s_{1,2}$, as well as their corresponding
ghosts.\footnote{%
At this stage we can also let the valence and sea quark masses differ.}
The sea-quarks are as in QCD. 
$O_2$ then matches onto
\begin{equation}
O_2^{PQA} = 2\left\{
[\bar s_1^a (1-\gamma_5) d_1^a][\bar s_2^b (1-\gamma_5) d_2^b]
+
[\bar s_1^a (1-\gamma_5) d_2^a][\bar s_2^b (1-\gamma_5) d_1^b]
\right\}
\,.
\end{equation}
The two terms correspond to the two types of Wick contraction in
QCD---note the $[\bar s_1 d_2][\bar s_2 d_1]$ flavor structure in the
second term.
The overall factor of 2 compensates for an overall
reduction by 2 in the number of Wick contractions.
This matching is exact to all orders in PT.
The point of this step is to separate the two Wick contractions,
without yet having to deal with any staggered complications.

In {\bf step 2} we stay in the same PQ continuum theory, but
match onto operators having $[\bar s_1 d_1][\bar s_2 d_2]$ form.
This is in preparation for matching to the staggered theory.
The result is a linear combination of four operators (note the
different Dirac and color structures):
\begin{eqnarray}
\lefteqn{O_2^{PQB} = 
b_1 [\bar s_1^a (1-\gamma_5) d_1^a][\bar s_2^b (1-\gamma_5) d_2^b]
+
b_2 [\bar s_1^a (1-\gamma_5) d_1^b][\bar s_2^b (1-\gamma_5) d_2^a]}
\\
&+&
b_3 [\bar s_1^a \sigma_{\mu\nu} (1-\gamma_5) d_1^a]
    [\bar s_2^b \sigma_{\mu\nu} (1-\gamma_5) d_2^b]
+
b_4 [\bar s_1^a \sigma_{\mu\nu} (1-\gamma_5) d_1^b]
    [\bar s_2^b \sigma_{\mu\nu} (1-\gamma_5) d_2^a]\,.
\label{eq:step2}
\end{eqnarray}
The coefficients $b_j$ have the form $b_j^{(0)}+\alpha b_j^{(1)}+ \dots$
and the LO and NLO coefficents are known. The $b_j^{(0)}$
can be obtained by a Fierz transform, while the $b_j^{(1)}$ require
a 1-loop calculation and one must make a choice of evanescent operators.

In {\bf step 3} we match onto the ``staggered partially quenched''
(SPQ) continuum theory, which contains sea quarks
$U$, $D$ and $S$  and valence quarks $D_{1,2}$ and $S_{1,2}$
(with uppercase again indicating four tastes)
as well as sufficient ghosts (25 in all) to
cancel both the valence determinant and the contribution of 3 of the
4 tastes for each sea quark.
Here, for simplicity, we drop the negative parity parts of the operator,
which do not contribute to the matrix element we are ultimately interested in.
The operator matching is exemplified by
\begin{equation}
[\bar s_1^a (1\!-\!\gamma_5) d_1^a][\bar s_2^b (1\!-\!\gamma_5) d_2^b]
\longrightarrow
\frac1{N_t} \left\{
[\bar S_1^a (1\otimes\xi_5) D_1^a][\bar S_2^b (1\otimes\xi_5) D_2^b]
+
[\bar S_1^a (\gamma_5\otimes\xi_5) D_1^a]
[\bar S_2^b (\gamma_5\otimes\xi_5) D_2^b]
\right\}\,,
\end{equation}
where $N_t=4$ is the number of tastes. Here we are assuming external
kaons of taste $\xi_5$ created with normalized operators.
This matching is exact to all orders in PT.
Each operator in the sum (\ref{eq:step2}) is matched similarly
and the coefficients $b_j$ are unchanged.

We are now ready to match, in {\bf step 4},
 to a lattice theory with rooted staggered 
(asqtad) sea quarks $\chi_u$, $\chi_d$ and $\chi_s$ and 
(HYP) valence quarks $\chi_{d1,2}$ and $\chi_{s1,2}$.
Our continuum operator from step 3 is in the form that
is used in practice in staggered lattice calculations of
matrix elements. The matching from the continuum PQ theory
to the lattice theory is (at present) done at 1-loop in PT.
At this order in $\alpha$, a large number of lattice operators
contribute to the matching, although most have the wrong taste
and are dropped in numerical calculations.
The form of the lattice operators that are kept is
described in Ref.~\cite{bsmprd}. The errors
that are introduced at this stage have the form
\begin{equation}
O^{\rm LAT}_j \cong O^{\rm SPQ}_j + {\cal O}(\alpha) 
[\textrm{wrong taste ops}]
+{\cal O}(\alpha^2, a^2) [\textrm{various taste ops}]
\,,
\end{equation}
where $\cong$ indicates that matrix elements (rather than operators)
are being matched. 

In {\bf step 5} we match onto the Symanzik effective continuum theory
describing the long distance degrees of freedom of the lattice theory.
Lattice symmetries are used to determine corrections to the action and
operators, and factors of $a^2$ are now explicit.

Finally, in {\bf step 6} we match onto the chiral effective
theory (SChPT), using
the chiral transformation properties of the action and operators.
This process is straightforward, though tedious.
A crucial step in this matching (which we do not have space to enter into)
is that certain symmetry properties of the original operator percolate
through the multiple steps and constrain the low-energy constants
(LECs) that multiply the
chiral operators, reducing the number of independent constants.
The details of this are slightly modified from the discussion in
Ref.~\cite{bsmprd} by the addition of steps 1 and 2, but the final
conclusion is the same.

With the chiral form of the operators in hand, the requisite one-loop
calculation is now straightforward. We have done this in SU(3) SChPT,
and obtained the SU(2) result using the method explained and justified
in Refs.~\cite{bkprd,bsmprd}.

\section{Results and conclusions}

The general form of the NLO result in SChPT is 
\begin{equation}
B_j = B_j^{\rm LO} \left[1 + \delta B_j^{\rm anal}
+ \delta B_j^{\rm log}\right]
\,.
\end{equation}
The analytic terms involve five unknown LECs at NLO:
\begin{equation}
\delta B_j^{\rm anal} =
c_{j1} (m_x+m_y) + c_{j2} (m_u+m_d+m_s)
+c_{j3} a^2 + c_{j4} a_\alpha^2 + c_{j5} \alpha^2\,.
\end{equation}
Here we have shown the SU(3) result; for SU(2) the $m_y$ and $m_s$
terms are absent (and the coefficients are different).
The first two terms are present in the continuum, while the remaining
three represent lattice artifacts. The $a^2$ term comes from standard
discretization errors (which are not suppressed by $\alpha$ since our
valence fermions are not fully improved). The $a_\alpha^2=a^2 \alpha^2$ 
term comes from taste-breaking discretization errors,
which are numerically enhanced such that they are comparable to
standard ${\cal O}(a^2)$ errors. Finally, the
$\alpha^2$ term arises because we use 1-loop operator matching---it
would be absent if we used non-perturbative renormalization.

These analytic terms are similar to those for $B_K$
and one can deal with them in a similar fashion.
In a chiral extrapolation at fixed $a$ the last three terms are constants
and can be absorbed into $B_j^{\rm LO}$. These terms then enter into
a (fairly complicated) continuum extrapolation.

The non-analytic terms are completely predicted in terms of the
pion decay constant in the appropriate chiral limit ($f_3$ or $f_2$)
and the masses of the valence and sea pions.
They lead to curvature in the predicted chiral behavior.
We show explicitly only the SU(2) result
for $N_f=2+1$ (which is the most useful in practice): 
\begin{equation}
\delta B^{\rm log}_j = \pm \frac{1}{(4\pi f_2)^2}
\left[ -\frac1{16} \sum_B \ell(M^2_{xx;B})
+ \frac12 \left\{
\ell(M^2_{xx;I}+ (M^2_{\pi;I}-M^2_{xx;I}) \tilde\ell(M^2_{xx;I})\right\}
\right]\,.
\end{equation}
Here $M_{xx;B}$ is the LO mass of the valence $\bar x x$ pion of taste $B$
and $M_{\pi;I}$ is the LO mass of the taste-singlet sea-quark pion.
The chiral logarithms are $\ell(X) = X \ln(X/\mu_\chi^2)$
and $\tilde\ell(X) = - d\ell(X)/dX$. The $+$ sign holds for
$j=2$ and $3$, while the $-$ sign applies for $j=4$ and $5$.

Key features of this result are:
\begin{itemize}
\item
The chiral logarithms for $j=2$ and $3$ (and for $j=4$ and $5$) are identical
because the chiral transformation properties of these
pairs of operators are the same~\cite{damir_bsm}.
This holds generally---i.e. also for SU(3) SChPT, and irrespective of
degeneracies in sea-quark masses.
\item
The $j=2,3$ and $j=4,5$ cases differ only in sign. This 
relationship does {\em not} hold for the SU(3) result.
\item
The chiral log for $j=2,3$ is identical to that for $B_K$.
This identity does {\em not} hold for the SU(3) result.
\item
The result is unaffected by the use of a mixed action---this only
impacts some coefficients in $\delta B_j^{\rm anal}$ as well as the
values of the pion masses and decay constant.
This conclusion holds also for the SU(3) result (but not for $B_K$).
\end{itemize}

In Ref.~\cite{bsmprd} we present also the SU(2) result for non-degenerate
sea quarks, and the SU(3) result both for $N_f=1+1+1$ and $N_f=2+1$.
We only note here that the SU(3) chiral logarithms for the BSM operators are
much simpler than those for $B_K$. The latter involve 13 additional
LECs, whereas the former involve none.
This simplicity can be traced back to the fact that the matrix
elements of the BSM operators do not vanish in the SU(3) chiral limit.

Some time ago, Becirevic and Villadoro (BV) calculated the chiral 
logarithms for the
BSM operators in continuum, unquenched SU(3) ChPT~\cite{damir_bsm}
(results which we use as a  check on ours in the appropriate limit).
They noted that the chiral logarithmss 
cancel in the ratios $B_2/B_3$ and $B_4/B_5$. 
They dubbed these ratios ``golden'' and
advocated their use in minimizing errors in chiral extrapolations.
Our results allow us to extend the set of golden combinations.
First (and this is almost trivial) the two golden ratios of BV
remain golden in the PQ theory, and in SU(2) ChPT.
Second, the ratio $B_2/B_K$ becomes golden in SU(2) ChPT.
Finally, the product $B_2 B_4$ is also golden in SU(2) ChPT.
Neither of these new combinations is golden in SU(3) ChPT, 
although there is a partial cancellation of chiral logarithms 
(and so in BV's parlance they are ``silver'' combinations).

These golden combinations could be particularly important for a
calculation using staggered fermions. This potential importance
arises because chiral logarithms
are responsible for a significant fraction of the taste breaking 
in the $B_j$ (since the pion masses entering the logarithms are
taste dependent).
Thus the chiral behavior of the golden combinations
may be smoother and easier to fit. 
One need only fit one of the five B-parameters 
including chiral logarithms,
determining the other four using golden ratios or products.
We are presently testing this approach in our companion numerical 
calculation~\cite{stag_bsm_proc}.

\section*{Acknowledgments}
W.~Lee is supported by the Creative Research
Initiatives Program (2012-0000241) of the NRF grant funded by the
Korean government (MEST), and acknowledges
support from KISTI supercomputing
center through the strategic support program for the supercomputing
application research [No. KSC-2011-G2-06].
S.~Sharpe is supported in part by the US DOE grant
no.~DE-FG02-96ER40956.


\begin{thebibliography}{99}
  \bibitem{dwf_bsm_pap}
  P.~A.~Boyle, N.~Garron and R.~J.~Hudspith,
\emph{Neutral kaon mixing beyond the standard model with $n_f = 2+1$ 
      chiral fermions,}
  Phys.\ Rev.\ D {\bf 86}, 054028 (2012)
  [{\tt arXiv:1206.5737 [hep-lat]}].

  \bibitem{tm_bsm_pap}
  V.~Bertone {\em et al.},
\emph{Kaon Mixing Beyond the SM from Nf=2 tmQCD and model independent 
constraints from the UTA,}
  {\tt arXiv:1207.1287 [hep-lat].}

  \bibitem{stag_bsm_proc} J.A.~Bailey {\em et al.}, SWME Collaboration,
  these proceedings:
\emph{Beyond the Standard Model corrections to $K^0-\bar{K}^0$ mixing,}
{\tt arXiv:1211.1101 [hep-lat].} 

  \bibitem{mescia12}  
  F.~Mescia and J.~Virto,
\emph{Natural SUSY and Kaon Mixing in view of recent results from Lattice QCD,}
  {\tt arXiv:1208.0534 [hep-ph]}.

  \bibitem{bsmprd} 
 J.~A.~Bailey, H.~-J.~Kim, W.~Lee and S.~R.~Sharpe,
  \emph{Kaon mixing matrix elements from beyond-the-Standard-Model operators 
  in staggered chiral perturbation theory}
  Phys.\ Rev.\ D {\bf 85}, 074507 (2012)
  [{\tt arXiv:1202.1570 [hep-lat]}].

  \bibitem{buras}
  A.~J.~Buras, M.~Misiak and J.~Urban,
\emph{Two loop QCD anomalous dimensions of flavor changing 
  four quark operators within and beyond the standard model,}
  Nucl.\ Phys.\ B {\bf 586}, 397 (2000)
  [{\tt hep-ph/0005183}].

  \bibitem{Kim:2011pz}
  J.~Kim, W.~Lee and S.~R.~Sharpe,
  \emph{One-loop matching of improved four-fermion staggered operators with 
  an improved gluon action,}
  Phys.\ Rev.\ D {\bf 83}, 094503 (2011)
  [{\tt arXiv:1102.1774 [hep-lat]}].

  \bibitem{VdWSS}
  R.~S.~Van de Water and S.~R.~Sharpe,
\emph{$B_K$ in staggered chiral perturbation theory,}
  Phys.\ Rev.\ D {\bf 73}, 014003 (2006)
  [{\tt hep-lat/0507012}].

  \bibitem{bkprd}
  T.~Bae, {\em et al.},
\emph{$B_K$ using HYP-smeared staggered fermions in $N_f=2+1$ unquenched QCD,}
  Phys.\ Rev.\ D {\bf 82}, 114509 (2010)
  [{\tt arXiv:1008.5179 [hep-lat]}].

  \bibitem{LS_SChPT}
  W.~-J.~Lee and S.~R.~Sharpe,
\emph{Partial flavor symmetry restoration for chiral staggered fermions,}
  Phys.\ Rev.\ D {\bf 60}, 114503 (1999)
  [{\tt hep-lat/9905023}].

  \bibitem{AB_SChPT}
  C.~Aubin and C.~Bernard,
\emph{Pion and kaon masses in staggered chiral perturbation theory,}
  Phys.\ Rev.\ D {\bf 68}, 034014 (2003)
  [{\tt hep-lat/0304014}].

  \bibitem{damir_bsm}
  D.~Becirevic and G.~Villadoro,
\emph{Remarks on the hadronic matrix elements relevant to the SUSY 
$K_0- \overline{K_0}$ mixing amplitude,}
  Phys.\ Rev.\ D {\bf 70}, 094036 (2004)
  [{\tt hep-lat/0408029}].

  \bibitem{inprogress}
  J.A.~Bailey, H.-J.~Kim, W.~Lee and S.R.~Sharpe, in progress.

\end{thebibliography}
\end{document}